\documentclass[manuscript]{acmart}

\AtBeginDocument{%
  }

\setcopyright{acmlicensed}
\copyrightyear{2018}
\acmYear{2018}
\acmDOI{XXXXXXX.XXXXXXX}

\acmConference[MobileHCI '24 Extended Abstracts]{26th International Conference on Human-Computer Interaction with Mobile Devices and Services}{Sep 30 -- Oct 03,
  2024}{Melbourne, Australia}
\acmISBN{978-1-4503-XXXX-X/18/06}

\begin{document}

\title{The Atlas of AI Incidents in Mobile Computing: Visualizing the Risks and Benefits of AI Gone Mobile}

\author{Edyta Bogucka}
\affiliation{%
  \institution{Nokia Bell Labs}
  \city{Cambridge}
  \country{United Kingdom}
}
\email{edyta.bogucka@nokia-bell-labs.com}

\author{Marios Constantinides}
\affiliation{%
  \institution{Nokia Bell Labs}
  \city{Cambridge}
  \country{United Kingdom}
}
\email{marios.constantinides@nokia-bell-labs.com}

\author{Julia De Miguel Velazquez}
\affiliation{%
  \institution{Kings College London}
  \city{London}
  \country{United Kingdom}
}
\email{julia.de_miguel_velazquez@kcl.ac.uk}

\author{Sanja Šćepanović}
\affiliation{%
  \institution{Nokia Bell Labs}
  \city{Cambridge}
  \country{United Kingdom}
}
\email{sanja.scepanovic@nokia-bell-labs.com}

\author{Daniele Quercia}
\affiliation{%
  \institution{Nokia Bell Labs}
  \city{Cambridge}
  \country{United Kingdom}
}
\email{daniele.quercia@nokia-bell-labs.com}

\author{Andrés Gvirtz}
\affiliation{%
  \institution{Kings College London}
  \city{London}
  \country{United Kingdom}
}
\email{andres.gvirtz@kcl.ac.uk}

\renewcommand{\shortauthors}{Bogucka et al.}

\begin{abstract}
Today's visualization tools for conveying the risks and benefits of AI technologies are largely tailored for those with technical expertise. To bridge this gap, we have developed a visualization that employs narrative patterns and interactive elements, enabling the broader public to gradually grasp the diverse risks and benefits associated with AI. Using a dataset of 54 real-world incidents involving AI in mobile computing, we examined design choices that enhance public understanding and provoke reflection on how certain AI applications—even those deemed low-risk by law—can still lead to significant incidents. \\\texttt{Visualization:} \url{https://social-dynamics.net/mobile-ai-risks}
\end{abstract}




\maketitle

\section{Introduction}
Artificial Intelligence (AI) is now omnipresent, seamlessly integrating into mobile devices and profoundly influencing sectors like workplaces, healthcare, and education. These sensor-equipped, AI-powered devices have transformed workplace productivity by enabling effortless communication and efficient task management~\cite{mirjafari2019differentiating, aseniero2020meetcues, choi2021kairos, constantinides2020comfeel}. They also provide cost-effective, objective metrics on physical and mental health, facilitating large-scale behavioral monitoring~\cite{zhou2023circadian, park2023social, constantinides2018personalized, perez2021wearables}. In education, real-time data from these devices assist students in planning and controlling their learning activities~\cite{ciolacu2019education}. However, the rise of AI in mobile devices also brings substantial risks, particularly in terms of privacy and security. AI can lead to data leakage and unauthorized surveillance, raising serious concerns about the confidentiality and integrity of personal information. Advances in federated learning and differential privacy offer promising ways to mitigate these risks~\cite{mcmahan2017communication, dwork2008differential}, but ensuring transparency in AI decision-making remains a formidable challenge, especially in sensitive fields like healthcare. Moreover, security issues extend beyond data breaches to include physical risks, such as skin irritations caused by device materials~\cite{mills2016wearing, fitbit_irritations}. Thus, while AI-enhanced mobile technologies offer significant benefits, they demand rigorous risk assessments and public engagement to fully understand their implications \cite{constantinides2024_risks_benefits}.

Current tools for communicating the risks and benefits of AI technologies are predominantly aimed at technically proficient audiences. These tools, including impact assessment reports~\cite{microsoft2022Assessment, stahl2023systematicReview}, harm description templates~\cite{buccinca2023aha}, and databases such as the IBM Risk Atlas~\cite{IBMriskAtlas} and AI Risk Database~\cite{AIRiskDatabase}, focus on detailing technical risks for AI practitioners~\cite{constantinides2023prompts, interpretabilityMLRoles}, often neglecting the broader societal implications. Out of nearly four hundred AI auditing tools, only two attempted to communicate risks to non-specialists~\cite{ojewale2024towards}, revealing a significant communication gap. This gap leaves the general public without a comprehensive understanding of AI's potential trade-offs and broader impacts. To bridge this gap, we designed a visualization that uses narrative patterns and interactive elements to make the risks and benefits of AI in mobile computing accessible to a wide audience. We populated this visualization with data from 54 real-world AI incidents reported in the news, exploring design choices that enhance public understanding and provoke reflection on how even low-risk AI uses can lead to significant incidents.

\section{Related Work}
\label{sec:related}
We next reviewed research across three  areas: the risks and benefits of AI in mobile uses, databases of AI incidents, and visualizations for technology risks and benefits.
\smallskip

\noindent\textbf{Risk and Benefits of AI in Mobile Computing.} Integrating AI into mobile devices offers substantial benefits but also significant risks. These risks include data leakage and unauthorized surveillance, intensified by the intimate nature of mobile devices~\cite{haris2014privacy, constantinides2022good, das2023algorithmic}. Advances in federated learning and differential privacy are helping mitigate these risks by enhancing data security and allowing AI models to be trained directly on devices, thus preventing exposure of sensitive data~\cite{mcmahan2017communication, dwork2008differential}. Despite these improvements, challenges in AI explainability persist, particularly in high-stakes areas such as healthcare, where transparency is paramount~\cite{saraswat2022explainable}. On the benefit side, AI-powered mobile devices significantly enhance physical and mental health monitoring, facilitate large-scale behavioral monitoring~\cite{zhou2023circadian, park2023social, constantinides2018personalized, perez2021wearables}, and enable real-time data collection and analytics in educational settings~\cite{ciolacu2019education}.
\smallskip

\noindent\textbf{Databases of AI Incidents.} Databases cataloging AI incidents are crucial for understanding and mitigating the socio-technical risks associated with AI. These databases gather information from various sources, including internet news reports and contributions from AI communities. The AI Incident Database, for example, automatically collects online news reports of AI-related harms~\cite{mcgregor2021preventing}. The OECD AI Incidents Monitor tracks international media sources to document failures in human-AI interactions~\cite{oecdMonitor}. Meanwhile, the AI, Algorithmic, and Automation Incidents and Controversies Repository takes a curated approach, involving human oversight to classify and review incidents~\cite{aiaaic}. Technical databases like the Adversarial Threat Landscape for Artificial-Intelligence Systems focus on adversarial techniques and real-world attack case studies~\cite{mitreatlas}. The forthcoming AI Vulnerability Database aims to support developers and auditors by cataloging instances of AI failures~\cite{avidml}. These databases often provide publicly accessible data~\cite{aiaaic}  and use detailed taxonomies~\cite{oecdMonitor,mcgregor2021preventing}, enabling the analysis of incident patterns and effective communication to various audiences~\cite{wei2022ai}.
\smallskip

\clearpage
\noindent\textbf{Visualizations for Technology Risks and Benefits.} Visualization tools play a key role in elucidating how AI technologies operate, uncovering biases in datasets, and highlighting model vulnerabilities~\cite{vis4ml_review2024}. For example, visualizing disparities within datasets can reveal sampling biases, while comparing model inputs and outputs can expose decision-making inconsistencies~\cite{aif360-oct-2018, IBMriskAtlas}. Visualizations that help internalize model logic, assess risk, and calibrate trust can support many different types of stakeholders, particularly model makers (e.g., developers, data scientists), model breakers (e.g., product managers, risk officers, model adopters), and model users (e.g., domain experts or laypeople)~\cite{interpretabilityMLRoles}.
\smallskip

\noindent\textbf{Research Gap.} Current visualizations often target technically proficient audiences familiar with complex concepts like confusion matrices or activation maps~\cite{vis4ml_review2024}. To communicate AI risks and benefits effectively to non-experts, information needs to be simplified and made relatable to personal contexts~\cite{visWhatWorks2021}. This can be achieved through visualization techniques that incorporate narratives~\cite{narrativeViz}, metaphors~\cite{tacitMetaphors, data_hazards}, interactive elements~\cite{visxAI}, and personalized presentations ~\cite{climateChange}, making the risks and benefits of AI more accessible and engaging for the general public.

\section{Visualizing the Risks and Benefits of AI Uses}

\begin{figure*}[t!]
  \centering
  \includegraphics[width=\textwidth]{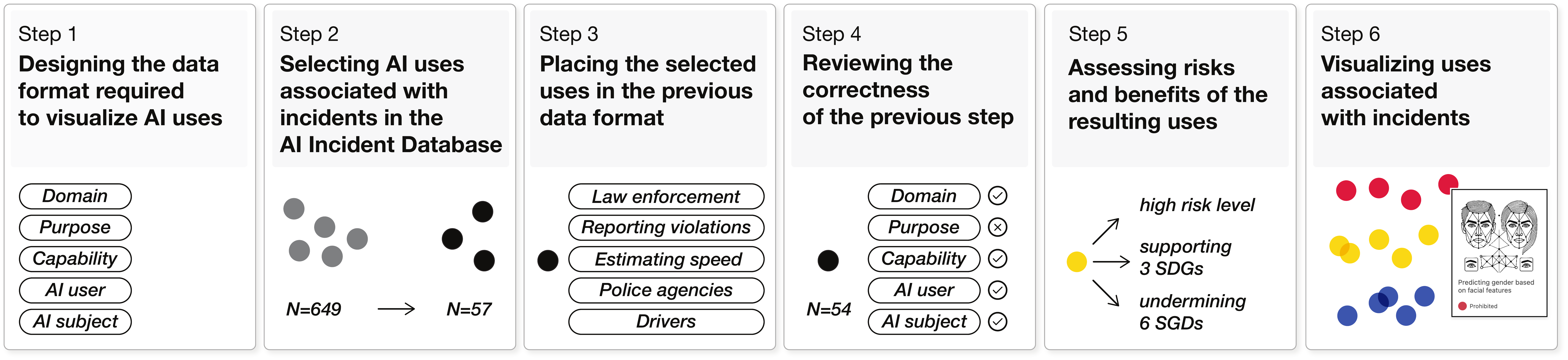}
  \caption{\textbf{Our six-step process for visualizing AI risks and benefits.} 
  We started with designing the data format required to visualize AI uses (Step 1). We then selected mobile computing uses associated with incidents in the AI Incident Database \cite{mcgregor2021preventing} (Step 2) and placed them in our data format (Step 3). Next, we reviewed the correctness of the use formatting (Step 4), manually assessed the risks and benefits of the resulting uses (Step 5), and finally, visualized these uses  (Step 6).}
  \label{fig:workflow}
\end{figure*}

To develop our visualization, we went through six steps (Figure~\ref{fig:workflow}). These steps included: establishing a data format to represent various AI uses; curating a dataset of 54 AI applications in mobile computing based on real-world incidents from the AI Incident Database (AIID); and developing a narrative-style visualization tool to highlight both the risks and benefits of these uses. We chose  AIID because it  provides a standardized and verified repository of AI incidents.

\subsection{Curating a Dataset of Risks and Benefits} 
\textbf{Step 1: Designing Data Format.} We created a standardized format to ensure consistent descriptions of AI applications across different technologies (first step in Figure \ref{fig:workflow}). This format, inspired by the EU AI Act's five-component risk assessment framework~\cite{Golpayegani2023Risk}, includes: the \emph{domain} specifying the application industry or sector (e.g., well-being), the \emph{purpose} explaining the goal (e.g., providing meditation guidance on a mobile app), the \emph{capability} describing the technology (e.g., real-time stress level tracking), the \emph{AI user} operating the system (e.g., health coaches), and the \emph{AI subject} affected by the system (e.g., app owners).

\clearpage
\vspace{1mm}
\noindent
\textbf{Step 2: Selecting AI Uses.} We focused on AI applications related to mobile computing (second step in Figure \ref{fig:workflow}), drawing from the AIID. By March 2024, the AIID contained 649 incidents, mostly from the US, spanning from 2013 to 2024. After eliminating duplicates, we analyzed 639 unique incidents, filtering those relevant to mobile devices. This process involved searching descriptions for mentions of mobile applications and technologies, ultimately selecting 57 uses.

\vspace{1mm}
\noindent
\textbf{Step 3: Formatting the Data.} Using OpenAI's GPT-4.0 API~\cite{OpenAI}, we paraphrased incident descriptions into detailed accounts of the AI uses involved (third step in Figure \ref{fig:workflow}). These were then structured into our predefined format. For instance, incident \#264, related to an AI system that measures vehicle speeds using a mobile phone camera~\cite{incidentSpeedcamAnywhere}, was categorized as: \emph{[Domain: Law enforcement, Purpose: Documenting and reporting traffic violations from video data, Capability: Estimating vehicle speed from video data, AI user: mobile app users, AI subject: drivers]}.

\vspace{1mm}
\noindent
\textbf{Step 4: Reviewing and Refining.}
Two authors who are experts in mobile computing reviewed the formatted uses, achieving 91\% agreement (fourth step in Figure \ref{fig:workflow}). Minor rephrasing was needed for five uses, and two duplicates were merged, resulting in a final set of 54 applications.

\begin{figure*}[t!]
  \centering
\includegraphics[width=\textwidth]{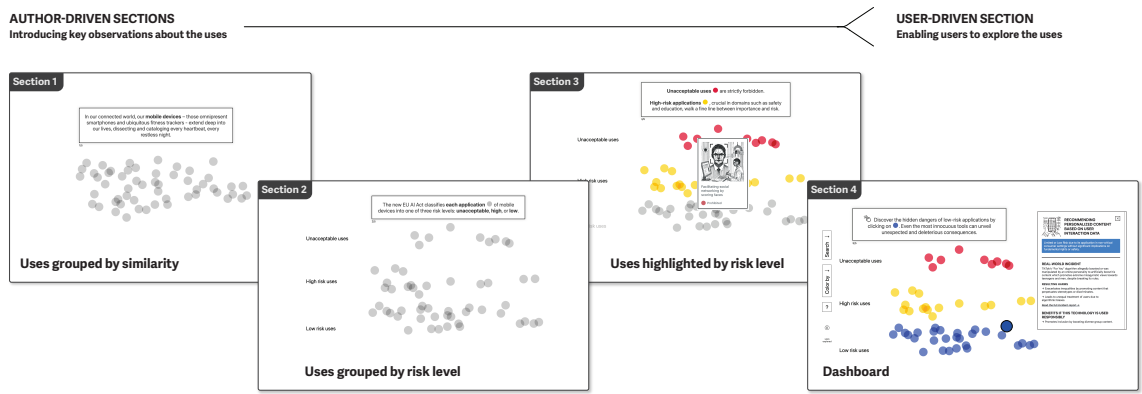}
  \caption{\textbf{The visualization of AI risks introduces information through the narrative structure of a Martini Glass (shown at the top) unfolding in four sections (shown at the bottom)}. The first three sections, author-driven, introduce the dataset of uses, risks, and benefits. The final section, user-driven, supports dataset exploration tasks through an interactive dashboard.}
  \label{fig:martini}
\end{figure*}

\vspace{1mm}
\noindent
\textbf{Step 5: Assessing Risks and Benefits.} We chose two frameworks to assess each use's risks and benefits: the EU AI Act~\cite{EUACT2024} and the 17 Sustainable Development Goals (SDGs)~\cite{sdgs}. The EU AI Act, the world's first AI regulation, follows a risk-based approach~\cite{EUACT2024}. It classifies the uses of technology based on the potential harm to society, dividing them into low-risk, high-risk, and unacceptable-risk categories, with stricter rules for higher risks. The SDGs address how technology use can impact people, planet, prosperity, peace, and partnerships~\cite{sdgs}, and can be used to assess AI's long-term impacts, both positive and negative~\cite{Cowls2021, constantinides2024_risks_benefits}.

With these two frameworks at hand, two experts in compliance classified the risks of each application according to the EU AI Act and evaluated their support for or potential to undermine the SDGs. The assessment identified 29 low-risk, 16 high-risk, and 9 unacceptable-risk uses, with a 90\% agreement between the experts. The uses were found to support nine SDGs while potentially undermining 14, with an 85\% agreement among the evaluators. For each use and SDG, the authors jointly listed one to three examples of potential risks and benefits. Discrepancies were resolved through discussions.

\subsection{Visualizing the Risks and Benefits}
\begin{figure*}[t!]
  \centering
\includegraphics[width=\textwidth]{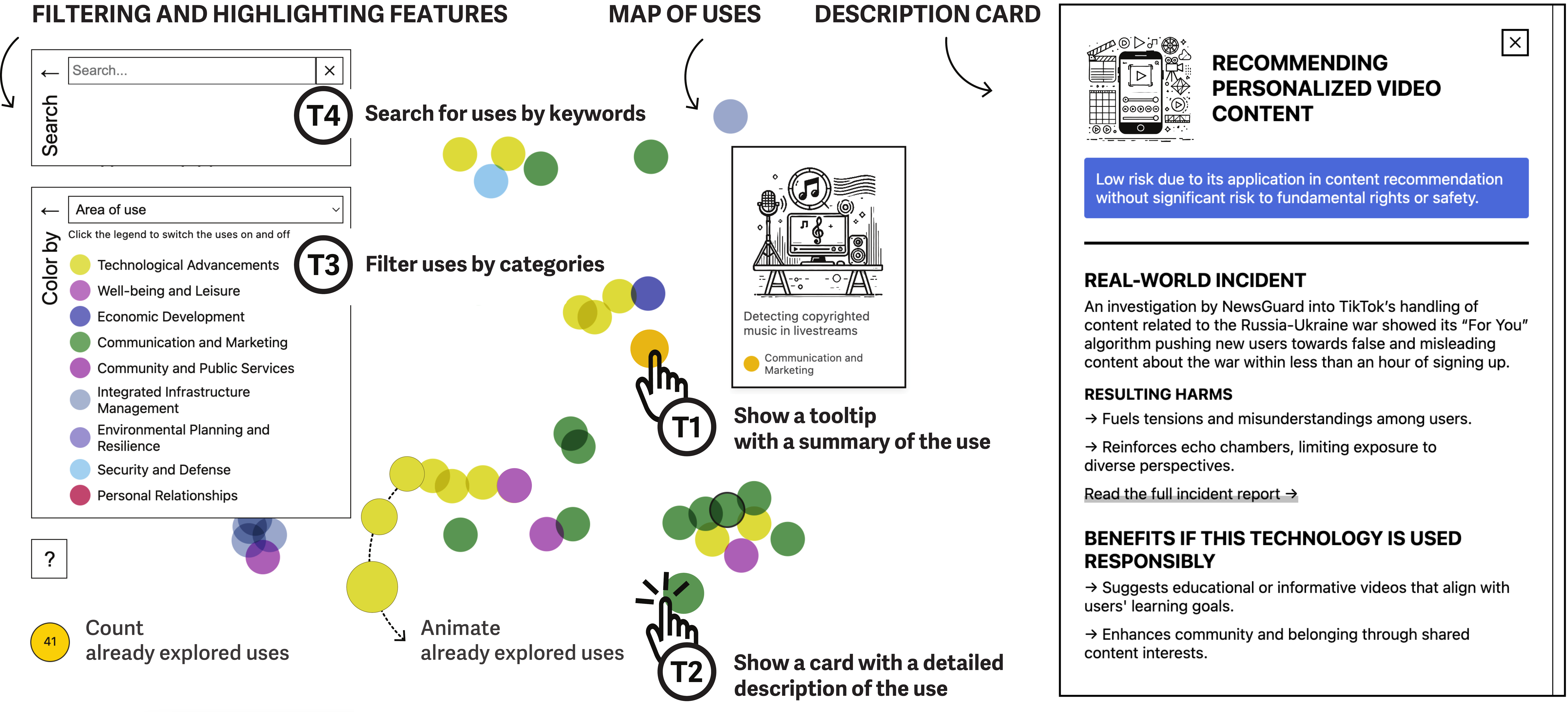}
  \caption{\textbf{The dashboard supports four dataset exploration tasks.} Users can browse the use map with tooltips (T1), view detailed descriptions of uses by clicking on dots (T2), filter them by categories (T3), and find similar uses through keyword search (T4).}
  \label{fig:narrative_tool}
\end{figure*}

To elucidate the risks and benefits associated with AI in mobile computing, we employed visual narrative design techniques (sixth step in Figure~\ref{fig:workflow}). These techniques simplify complexity and introduce information progressively (Figure~\ref{fig:martini}). The resulting tool (Figure~\ref{fig:narrative_tool}), a web-based application constructed with HTML5, JavaScript, and D3.js, is available online at~\url{https://social-dynamics.net/mobile-ai-risks}.

\smallskip
\noindent\textbf{Visual Narrative Design.} 
We structured the visualization using the Martini Glass model~\cite{narrativeViz}, which starts with guided storytelling before transitioning to user-driven exploration (Figure \ref{fig:martini}). This approach ensures that users are first introduced to key observations and then encouraged to delve deeper into the dataset on their own. The tool consists of four consecutive sections (Figure \ref{fig:martini}):

\begin{enumerate}

\item \textbf{Introduction to Mobile Computing Technologies:} The first section presents a map of AI uses in mobile computing, each represented as a dot~\cite{visWhatWorks2021}. We utilized sentence-level BERT (SBERT) embeddings~\cite{reimers2019sbert} and t-distributed stochastic neighbor embedding (t-SNE)~\cite{tsnePaper} to spatially distribute these dots based on semantic similarity.

\item  \textbf{Risk Levels:} The second section highlights how uses of the same technology can vary in risk, illustrated through an animated transition that categorizes the dots into unacceptable, high, or low risk groups.

\item  \textbf{Shared Traits of Risk Groups:} The third section explains the common characteristics of each risk category, emphasizing through color-coding how, despite regulations, all uses can still pose harm.

\item  \textbf{Interactive Dashboard:} The final section introduces a dashboard that encourages users to explore the low-risk uses, revealing that even these can cause harm if poorly implemented. For instance, TikTok's ``For You'' algorithm, though generally deemed low-risk, was linked to disinformation about the Ukraine war ~\cite{incidentTikTok}.

\end{enumerate}

\smallskip
\noindent\textbf{User Interface and Interaction Design.} The dashboard is divided into three main parts (Figure \ref{fig:narrative_tool}), corresponding to the so-called visual information-seeking mantra~\cite{mantra}:

\begin{enumerate}
\item \textbf{Overview:} The central map of uses allows users to hover over or click on dots to get summaries and risk tags (Figure \ref{fig:narrative_tool}, T1).

\item \textbf{Detailed Descriptions:} Clicking on a dot brings up a card with a comprehensive description, including real-world incident examples and potential benefits if the use is responsibly developed (Figure \ref{fig:narrative_tool}, T2).

\item \textbf{Filtering and Highlighting:} The left panel offers filtering options, allowing users to categorize uses based on properties such as application domain, AI subject, and AI user (Figure \ref{fig:narrative_tool}, T3). A keyword search box helps find similar uses, with results highlighted accordingly (Figure \ref{fig:narrative_tool}, T4).
\end{enumerate}

Our interface design employs bold and highly contrasting colors to distinguish between various use categories, ensuring they stand out prominently~\cite{visWhatWorks2021}. This design strategy also incorporates larger fonts and bold borders to enhance visibility, creating a clear contrast between interactive elements and the background.

\clearpage
\section{Discussion and Conclusion}
\label{sec:discussion}

Our visualization adopts a straightforward, five-component data format, effectively communicating the risks and benefits of specific AI applications, such as voice-activated mobile controls, and illustrating the broader impact of mobile computing technology. By representing uses as countable dots, grouped by similarities and distinguished by risk levels, we simplify complex AI relationships for a wide audience, including those with limited numeracy and AI literacy~\cite{visWhatWorks2021}.

The dual narrative approach—combining author-driven and user-driven elements—enables us to educate and empower diverse groups within the general public. For instance, in educational settings, our visualization can teach students about AI's ethical considerations and societal impacts~\cite{feffer2023ai}. In corporate environments, it can guide technology deployment strategies, ensuring employees understand both the advantages and potential hazards of AI~\cite{constantinides2022good}. In research contexts, it can showcase the community's contributions to the field by, for example, extracting uses from academic papers and creating conference dashboards~\cite{mappingConferences, chiPapers}.

Ultimately, our goal is to foster proactive discussions about mobile technology, weighing both its risks and benefits.

\bibliographystyle{ACM-Reference-Format}
\bibliography{sample-base}


\end{document}